\documentclass{aa}

\usepackage{graphicx}

\usepackage{txfonts}

\usepackage{natbib}
\bibpunct{(}{)}{;}{a}{}{,}

\begin{document}

\title{Spitzer/IRAC and ISOCAM/CVF insights on the origin of the Near
  to Mid-IR Galactic diffuse emission}

\subtitle{}

\author{N. Flagey
  \inst{1}
  \and F. Boulanger
  \inst{1} 
  \and L. Verstraete
  \inst{1} 
  \and M.A. Miville Desch{\^e}nes
  \inst{1}
  \and A. Noriega Crespo
  \inst{2}
  \and W.T. Reach
  \inst{2} 
}

\offprints{\\N. Flagey, \email{nicolas.flagey@ias.u-psud.fr}
}

\institute{
Institut d'Astrophysique Spatiale, Universit{\'e} Paris Sud, Bat. 121,
F-91405, Orsay Cedex, France\\
\and Spitzer Science Center, California Institute of Technology, 1200
East California Boulevard, MC 220-6, Pasadena, CA 91125
}

\date{Submitted to A\& A: August 1st, 2005}

\abstract{Spitzer/IRAC images of extended emission provide a new
  insight on the nature of small dust particles in the Galactic
  diffuse interstellar medium. We measure IRAC colors of extended
  emission in several fields covering a range of Galactic latitudes
  and longitudes outside of star forming regions. We determine the
  nature of the Galactic diffuse emission in Spitzer/IRAC images by
  combining them with spectroscopic data. We show that PAH features
  make the emission in the IRAC 5.8 and 8.0 $\mu$m channels, whereas
  the 3.3 $\mu$m feature represents only 20 to 50\% of the IRAC 3.6
  $\mu$m channel. A NIR continuum is necessary to account for IRAC 4.5
  $\mu$m emission and the remaining fraction of the IRAC 3.6 $\mu$m
  emission. This continuum cannot be accounted by scattered light. It
  represents 9\% of the total power absorbed by PAHs and 120\% of the
  interstellar UV photon flux. The 3.3 $\mu$m feature is observed to
  vary from field-to-field with respect to the IRAC 8.0 $\mu$m
  channel. The continuum and 3.3 $\mu$m feature intensities are not
  correlated.

  We present model calculations which relate our measurements of the
  PAHs spectral energy distribution to the particles size and
  ionization state. Cation and neutral PAHs emission properties are
  inferred empirically from \object{NGC7023} observations. PAHs
  caracteristics are best constrained in a line of sight towards the
  inner Galaxy, dominated by the Cold Neutral Medium phase : we find
  that the PAH cation fraction is about 50\% and that their mean size
  is about 60 carbon atoms. A significant field-to-field dispersion in
  the PAH mean size, from 40 to 80 carbon atoms, is necessary to
  account for the observed variations in the 3.3 $\mu$m feature
  intensity relative to the IRAC 8.0 $\mu$m flux. However, one cannot
  be secure about the feature interpretation as long as the continuum
  origin remains unclear. The continuum and 3.3 $\mu$m feature
  emission process could be the same even if they do not share
  carriers.

}

\titlerunning{Origin of the NIR to MIR diffuse emission}

\maketitle

\section{Introduction}

Interstellar emission in the near infrared (NIR) traces the properties
of the smallest dust particles known as polycyclic aromatic
hydrocarbons (PAHs). Their presence in the diffuse interstellar medium
(ISM) was inferred from photometric measurements with the {\it Infra
Red Astronomical Satellite} (IRAS) \citep{Boulanger1985}, AROME
\citep{Giard1994} and the {\it Diffuse Infra Red Background
Experiment} (DIRBE) \citep{Dwek1997}. First spectroscopic evidence was
provided by the {\it Infra Red Telescope in Space} (IRTS)
\citep{Tanaka1996} and the {\it Infrared Space Observatory} (ISO)
\citep{Mattila1996} for the Galactic plane. ISO succeeded to detect
the PAHs bands in spectra of high latitude cirrus clouds for $\lambda
> 5\ \mu$m \citep{Boulanger2000}. Measurements of the shorter
wavelength emission were still limited to bright objects such as
visual reflection nebulae \citep{Verstraete2001,Van2004}. For the
first time, with the {\it Infra Red Array Camera} (IRAC) on board {\it
Spitzer Space Telescope} (SST), the sensitivity and angular resolution
are available to measure the NIR interstellar emission independently
of stellar emission modeling, unlike with DIRBE. First determination
of IRAC colors in Galactic fields was reported in the first round of
Spitzer publications by \citet{Lu2004} from total power sky
brightnesses. We undertake a more thorough study to quantify what can
be learned on PAHs and the smallest dust particles with IRAC images of
the diffuse Galactic emission.

Within PAH emission models, measurement of the 3.3 $\mu$m feature is
critical to constrain the PAH ionization state and size
\citep{Li2001}. The existence of PAHs with a few tens of atoms was
proposed to account for the 3.3 $\mu$m emission
\citep{Leger1984}. Moreover, a continuum underlying the 3.3 $\mu$m
feature has been detected in visual reflection nebulae
\citep{Sellgren1983}, and more recently in galaxies
\citep{Lu2003}. This continuum is not accounted for in PAH models and
its origin is still open : is it fluorescence emission from PAHs or
photoluminescence from larger grains ?

In section 2, we present the IRAC data taken from the Galactic First
Look Survey (GFLS) and Galactic Legacy Infrared Mid-Plane Survey
Extraordinaire (GLIMPSE) as well as complementary ISOCAM/CVF
spectra. We measure IRAC colors of the Galactic diffuse emission
(section 3) and combine them with spectroscopic data in section 4 to
separate the contributions of the 3.3 $\mu$m feature and the continuum
to IRAC 3.6 $\mu$m channel. In section 5, we use an updated version of
the model of \citet{Desert1990} -- detailed in the appendix -- to
bring constraints on the PAH mean size and ionization state (section
6). We discuss the origin of the NIR continuum in section 7.

\begin{figure}
\resizebox{\hsize}{!}{\includegraphics{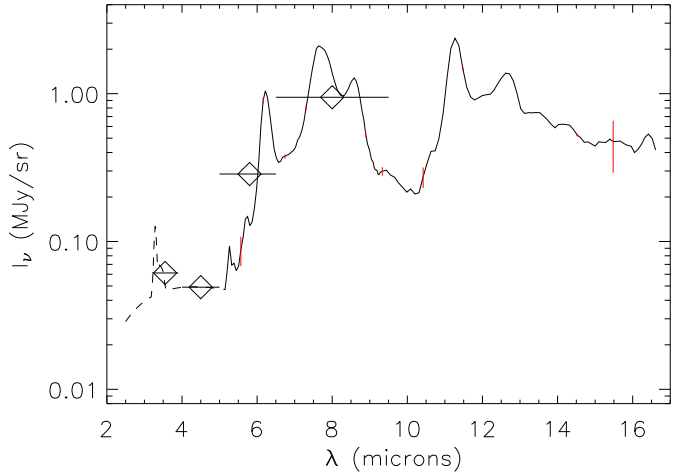}}
\caption{\textit{Solid line} : CVF spectrum ($\lambda / \delta \lambda
  = 35 \rightarrow 45$) of the diffuse Galactic emission (for
  $N_\mathrm{H}=10^{21} \mathrm{cm}^{-2}$) centered on the Galactic
  coordinates (26.8,+0.8), completed for $\lambda < 5 \mu$m with the
  IRAC colors measured on the GLIMPSE field (\textit{diamonds}).
  Horizontal lines represent IRAC filter widths. \textit{Dashed line}
  : the gray body, with a color temperature $T_C = 1105 \mathrm{K}$
  and the 3.3 $\mu$m feature, whose shape is extracted from the SWS
  spectrum of \object{NGC7023} and intensity is the one measured by
  \citet{Giard1994}. Vertical lines represents some error bars on the
  spectrum.}
\label{fig:spectre}
\end{figure}

\section{Observations}
\label{sec:obs}

\subsection{Selected fields}
For this study, we use images from the GFLS and one field from GLIMPSE
that span a range of Galactic longitudes and latitudes. These fields
point towards the diffuse Galactic medium, away from bright star
forming regions over path lengths which increase with decreasing
Galactic longitudes and latitudes. The GLIMPSE field is centered on
Galactic plane at a longitude $l$ = 27.5$\degr$ and extends over an
area of 3$\degr$ by 20$\arcmin$. It is a mosaic of 71x4 fields of
5$\arcmin$ by 5$\arcmin$ with an individual exposure time of 2
seconds. The GFLS fields are centered on Galactic coordinates ($l,b$)
= (254.4,+0), (105.6,+0.3), (105.6,+4), (105.6,+8), (105.6,+16) and
(105.6,+32) and cover an area of 1$\degr$ by 15$\arcmin$. They are
mosaics of 3x12 fields of 5$\arcmin$ by 5$\arcmin$ with an individual
exposure time of 12 seconds.

We complement IRAC fields with ISOCAM/CVF spectroscopic data covering
the 5 to 16 $\mu$m wavelength range at the positions listed in Table
\ref{tab:r7_11} over a 3$\arcmin$ by 3$\arcmin$ area. These
observations also point towards the diffuse Galactic medium, away from
bright star forming regions. There is only one position common to IRAC
and ISOCAM/CVF, centered on Galactic coordinates (26.8,+0.8). We
estimate the gas column density along this line of sight from the
$\mathrm{H~I}$ Leiden/Dwingeloo survey \citep{Burton1994} and the
Columbia $\mathrm{CO}$ survey \citep{Cohen1986}. The total column
density is $N_\mathrm{H}=2.10^{22}\ \mathrm{cm}^{-2}$ including
$5.10^{21}\ \mathrm{H_{2}\ cm}^{-2}$. For $N_\mathrm{H_{I}}$ we
assume that the emission is optically thin, and for $N_\mathrm{H_2}$ we
use the conversion factor $2.8\ 10^{20}\ \mathrm{H_2}\
\mathrm{cm}^{-2}$ per unit $\mathrm{CO}$ emission expressed in
$\mathrm{Kkms}^{-1}$.

\subsection{ISOCAM/CVF spectroscopy}
\label{sec:cvf}

\begin{figure}[!t]
\resizebox{\hsize}{!}{\includegraphics{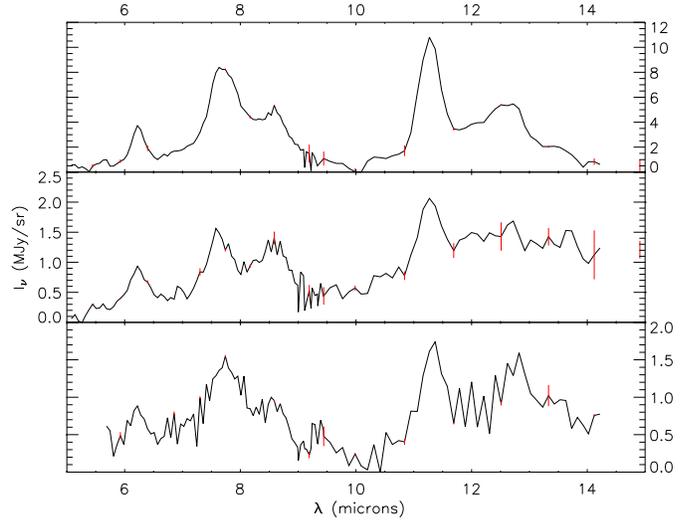}}
\caption{CVF spectra ($\lambda / \delta \lambda = 35 \rightarrow
  45$) of the diffuse Galactic emission centered on the Galactic
  coordinates (34.1,13.4), (299.7,-16.3) and (30.0,3.0), from bottom
  to top. Vertical lines represents some errors bars on the spectra.}
\label{fig:spectres_cvf}
\end{figure}

The ISOCAM/CVF spectra are taken from the ISO Archive. We use the
Highly Processed Data Products (HPDP) as described by
\citet{Boulanger2005}. The spectral resolution is between 35 and 45
for $\lambda$ between 5 and 16.5 $\mu$m. We produce a mean spectrum by
averaging all spectra over the 3$\arcmin$ by 3$\arcmin$ field of
view. In each spectrum the PAH features are visible. The highest S/N
spectrum, that centered on Galactic coordinates (26.8,+0.8), is shown
on Fig.~\ref{fig:spectre}. The three others are plotted on
Fig.~\ref{fig:spectres_cvf}. Error bars on the spectra, represented by
vertical lines, were obtained by comparing spectra computed over
distinct sub-areas. They are conservative estimates of the error bars
as they may include true variations in the sky emission. The
uncertainties are dominated by systematic effects (detector
transients, zodiacal light subtraction) and are correlated over
wavelengths \citep{Boulanger2005}.

\subsection{IRAC images and processing}
\label{images}

\begin{figure}
\resizebox{\hsize}{!}{\includegraphics{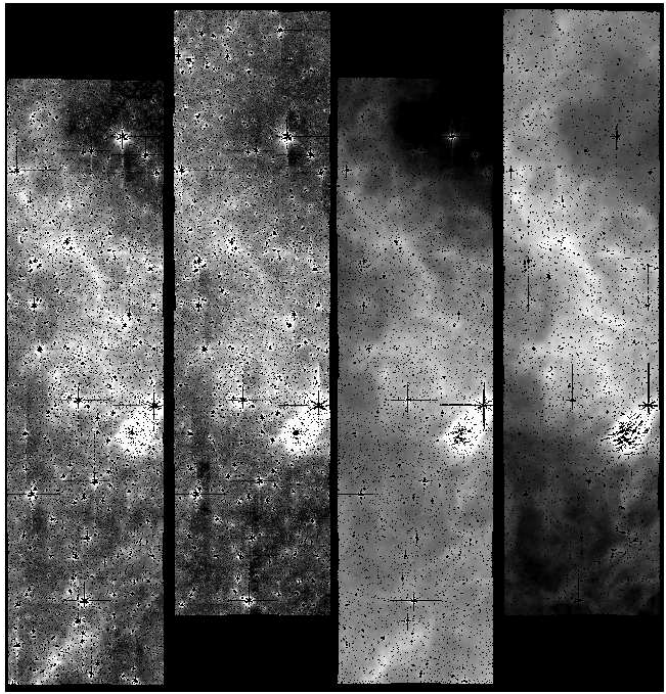}}
\caption{IRAC images (1$\degr$ by 15$\arcmin$) centered on Galactic
  coordinates (105.6,+4) as viewed in 3.6, 4.5, 5.8 and 8.0 $\mu$m
  channels (\textit{from left to right}). Most of the point sources
  are masked (black dots on the images) and a median filter is
  applied. The diffuse Galactic emission is clearly visible in each
  channel. Color scales are different from one channel to
  another : the brightest areas are in white (0.4, 0.3, 4 and 10
  MJy/sr for 3.6, 4.5, 5.8 and 8.0 $\mu$m channels), the faintest in
  black (0.1, 0.05, 1 and 3 MJy/sr).}
\label{fig:irac}
\end{figure}

All the IRAC images come from the Spitzer Archive \footnote{See
http://ssc.spitzer.caltech.edu/archanaly/status/} (pipeline software
version S11.0.2). We use the mosaiced images (post-bcd). The diffuse
emission is clearly visible in almost all IRAC data at a Galactic
latitude below 16$\degr$ (see Fig.~\ref{fig:irac}) but not at
(105.6,+32) which is located in a low column density region and that
we use as an estimator of the noise. The surface brightness
sensitivities are 0.0397, 0.0451, 0.154 and 0.165 MJy/sr for IRAC 3.6,
4.5, 5.8 and 8.0 $\mu$m channels, as given by the Spitzer Sensitivity
Performance Estimation Tool for a low background level \footnote{See
http://ssc.spitzer.caltech.edu/tools/senspet/}. For the GLIMPSE field,
the integration time is shorter and the surface brightness
sensitivities are 0.409, 0.385, 1.03 and 0.886 MJy/sr for IRAC 3.6,
4.5, 5.8 and 8.0 $\mu$m channels, for a high background level.

We remove most of point sources by masking the pixels that are more
than 3$\sigma$ away from the image processed by a median filtering
window of 21x21 pixels (1 pixel = 1.2$\arcsec$). Since we study the
extended emission, we choose to apply a median filter (5x5 pixels) to
the four IRAC channels data and keep 1 pixel of every 3. This reduces
the image size and increase their signal-to-noise ratio. The effective
resolution of the images is then about 6 $\arcsec$. All images of a
given field are projected on a common grid. We finally apply the
photometric corrections given in Table 5.7 of the IRAC Data Handbook
\footnote{See http://ssc.spitzer.caltech.edu/irac/dh/} for ``infinite
aperture'' to all IRAC photometric results, since we focus on the
diffuse emission extended over a significant fraction of the
fields. After these corrections, the absolute calibration accuracy is
about 5\%.

\section{IRAC colors of the diffuse emission}
\label{irac}
We measure the diffuse emission colors on the GLIMPSE and GFLS fields
by correlating the intensity in two IRAC channels. We discuss
extinction correction and compute the ionized gas contribution to the
colors.

\subsection{Correlations}
\label{sec:corr}

\begin{figure}[t]
\resizebox{\hsize}{!}{\includegraphics{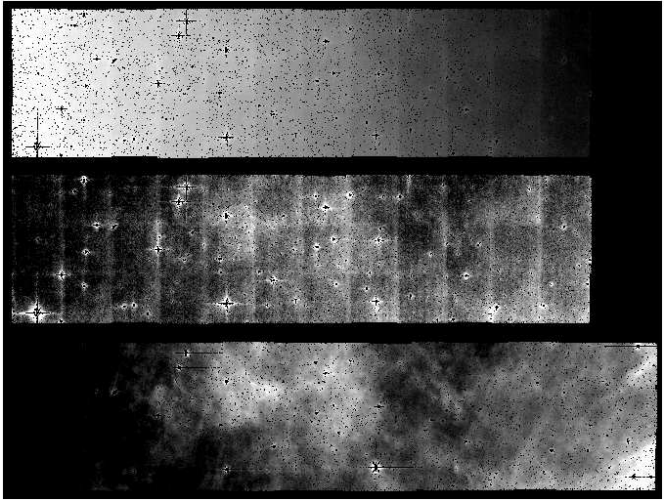}}
\caption{Field centered on Galactic coordinates (105.6,+8) as viewed
  by IRAC 5.8 $\mu$m channel before gradient correction
  (\textit{top}), by IRAC 5.8 $\mu$m channel after gradient correction
  (\textit{center}), and by IRAC 8.0 $\mu$m channel before gradient
  correction (\textit{bottom}). The small scale structures, visible on
  IRAC 8.0 $\mu$m channel before gradient correction exhibits on IRAC
  5.8 $\mu$m channel only after gradient correction.}
\label{fig:grad}
\end{figure}

\begin{figure}[t]
\resizebox{\hsize}{!}{\includegraphics{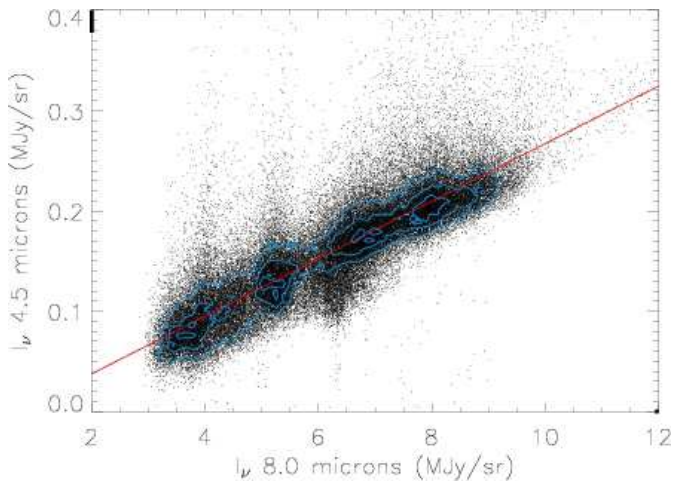}}
\caption{Correlation plot of IRAC 4.5 $\mu$m channel versus IRAC 8.0
  $\mu$m channel for the GFLS field centered on Galactic coordinates
  (105.6,+4). The curves are iso-density contours (from the inside to
  the outside, the density of plot is at least 75\%, 50\% and 25\% of
  the maximum density). The straight line is the result of the linear
  fitting. The statistical noise per pixel, as measured in the field
  at (l,b) = (105.6,+32), is 0.0268, 0.0222, 0.581 and 0.0639 MJy/sr
  for IRAC 3.6, 4.5, 5.8 and 8.0 $\mu$m channel.}
\label{fig:scat4}
\end{figure}

We measure the IRAC colors of the diffuse Galactic emission by
correlating the brightness structure in each IRAC channel with the
IRAC 8.0 $\mu$m channel (see Fig.~\ref{fig:scat4}). Some IRAC images
present a strong intensity gradient along the long axis, that is
obviously an artifact (see Fig.~\ref{fig:grad}), which appears
during the mosaicing process and is due to bad dark-current
correction. This gradient appears with different strengths on IRAC
channels. It generally seems to be weak on IRAC 8.0 $\mu$m channel,
relative to the Galactic diffuse emission. On the other IRAC channels,
especially at 5.8 $\mu$m, the emission structure is sometimes
dominated by a smooth gradient, that we take into account in the data
correlation. We thus decompose each IRAC 3.6, 4.5 and 5.8 $\mu$m
images into a gradient, represented by a low order polynomial function
of the long axis position, plus the emission structure of the IRAC 8.0
$\mu$m channel. The uncertainties on the IRAC colors, associated with
the gradient fitting, are estimated by looking at the variations of
the color ratios with the order of the gradient.

For each color, we iterate a linear regression, taking into account
the statistical noise per pixel as measured in the high latitude
field (0.0268, 0.0222, 0.581 and 0.0639 MJy/sr for IRAC 3.6, 4.5, 5.8
and 8.0 $\mu$m channel), and at each iteration, the pixels that are
more than 3$\sigma$ away from the linear fit are masked for the next
fit, where $\sigma$ is the standard deviation of the distance
between the points and the straight line. Such pixels correspond to
bright sources which are not removed by the median filter because they
extend over an area comparable to the filter window. Five iterations
are sufficient to converge.

\subsection{Extinction correction}
\label{sec:ext}

Total hydrogen column densities are estimated from HI and CO
observations (see section \ref{sec:obs}). These column densities are
converted into extinction in the IRAC channels combining $A_V/N_H =
0.53\ 10^{-21}\mathrm{cm}^{2}$ from \citet{Savage1979}, $A_K/A_V =
0.112$ \citep{Rieke1985} and the wavelength dependence of the
extinction in IRAC channels determined from stellar measurements with
GLIMPSE data \citep{Indebetouw2005}. We use their ``Average'' IRAC
extinctions normalized to $A_K$. The extinction corrections on the
IRAC colors are not negligible for the GLIMPSE field, where the gas
column density is the highest. For the other fields, the correction is
small compared to uncertainties.

For the GLIMPSE field, our $A_V$ is closer than 10\% to that given by
the maps of \citet{Schlegel1998} with $R_V=3.1$. We assume that the
emitting dust is mixed with the grains responsible for the
extinction. The following relation gives the intrinsic color
$R_{\lambda_{i}/\lambda_{j},int}$, ratio between $I_\nu$ in IRAC
$\lambda_{i}$ channel and $I_\nu$ in IRAC $\lambda_{j}$ channel, given
the extinction $\tau$ at both wavelengths and the observed color
$R_{\lambda_{i}/\lambda_{j},obs}$, measured in section \ref{sec:corr}
:

\begin{equation}
R_{\lambda_{i}/\lambda_{j},int} = R_{\lambda_{i}/\lambda_{j},obs}
  \times \frac{1-\exp(-\tau_j)}{1-\exp(-\tau_i)} \times \frac{\tau_i}{\tau_j}
\label{eq:ext}
\end{equation}

$R_{3.6/8.0}$ is the only color to which we apply this correction.
Differences in the extinction at 4.5, 5.8 and 8.0 $\mu$m are within
error bars ($\tau_{4.5} = \tau_{5.8} = \tau_{8.0} = 2.3\pm0.3$). The
3.6 $\mu$m opacity is significant ($\tau_{3.6} = 3.1$ at $b=0\degr$)
but the $R_{3.6/8.0}$ color correction is relatively small (1.24)
because the opacities at both wavelength are close to each other. The
extinction depends on the Galactic latitude. The $b=0\degr$ opacity
gives the relevant extinction correction since the GLIMPSE colors are
dominated by low-latitude emission.  In the data analysis we combine
IRAC colors with AROME 3.3 $\mu$m feature observations and CVF
spectroscopy. The extinction correction applied to the AROME
measurements is consistent with ours \citep{Giard1994}. The extinction
correction for the CVF field, located at $b=0.8\degr$, is negligible
for $\lambda > 5\ \mu$m.

\subsection{Ionized gas emission}
\label{sec:ff}

Free-free as well as gas lines emission might contribute to IRAC 3.6
and 4.5 $\mu$m channels, whatever the Galactic longitude. In order to
obtain the contribution of the free-free emission at 3.6 and 4.5
$\mu$m, we first measure this emission at 5 GHz. For the GLIMPSE
field, we obtain the variation of the free-free emission from
$b=1\degr$ to $b=0\degr$ at 5 GHz from \citet{Altenhoff1979}. For the
GFLS fields, we deduce the free-free emission at radio frequencies
from the $\mathrm{H_\alpha}$ emission \citep{Reynolds1992}. We measure
the $\mathrm{H_\alpha}$ emission on the $\mathrm{H_\alpha}$ Full Sky
Map corrected by extinction by \citet{Dickinson2003}. Then, we
extrapolate the electrons emission (free-free and free-bound) from 5
GHz to NIR according to \citet{Beckert2000}.

We finally add the contribution of gas lines. The fluxes from the main
H recombination lines within the IRAC channels (Pf$_\delta$ at
3.296$\mu$m, Pf$_\gamma$ at 3.739 $\mu$m, Br$_\alpha$ at 4.051 $\mu$m,
Pf$_\beta$ at 4.652$\mu$m, and Pf$_\alpha$ at 7.46$\mu$m ) per unit
Br$_\gamma$ emission are taken from the \citet{Hummer1987} Table for
an electron temperature and density of $7500\ \mathrm{K}$ and $10^2\
\mathrm{cm}^{-3}$, and case B recombination (nebula optically thick to
H ionizing photons). At 3.6 $\mu$m, they represents 20\% of the
free-free emission, whereas at 4.5 $\mu$m, they double its
contribution. The total contribution of the free-free and gas lines
emission is thus about 1\% at 3.6 $\mu$m and 3\% at 4.5 $\mu$m for
most of GFLS fields, except the (105.6,+8) field, for which the
contributions reach 3\% and 11\%. For the GLIMPSE field, the figures
deduced from \citet{Altenhoff1979} are about 7\% and 12\%.

\subsection{IRAC colors}

\begin{table*}[!t]
\caption{IRAC color ratios.}
\label{tab:rdc}
\centering
\renewcommand{\footnoterule}{}  % to avoid a line before footnotes
\begin{tabular}{l  r@{.}l@{ $\pm$ }r  r@{.}l@{ $\pm$ }r  r@{.}l@{
      $\pm$ }r r@{.}l@{ $\pm$ }r r@{.}l@{ $\pm$ }r c}

\hline \hline 

& \multicolumn{3}{c}{$R_{3.6/8.0}$} & \multicolumn{3}{c}{$R_{4.5/8.0}$}
& \multicolumn{3}{c}{$R_{5.8/8.0}$} & \multicolumn{3}{c}{$R_{3.6/8.0,
feat}$} & \multicolumn{3}{c}{$R_{3.6/8.0, cont}$} & $N_{\mathrm{C}}$ \\
\hline

Spitzer GFLS (105.6,+0.3) & 0&076 & $10\ 10^{-3}$ & 0&065 & $8\
10^{-3}$ & 0&37 & $5\ 10^{-2}$ & 0&014 & $2\ 10^{-3}$ $^{\mathrm{ a}}$
& 0&062 & $7\ 10^{-3}$ $^{\mathrm{ a}}$ & $80\pm20$ \\

Spitzer GFLS (105.6,+4) & 0&059 & $8\ 10^{-3}$ & 0&037 & $5\ 10^{-3}$
& 0&32 & $4\ 10^{-2}$ & 0&024 & $4\ 10^{-3}$ $^{\mathrm{ a}}$ & 0&035 & $5\
10^{-3}$ $^{\mathrm{ a}}$ & $56\pm12$ \\

Spitzer GFLS (105.6,+8) & 0&094 & $15\ 10^{-3}$ & 0&050 & $7\ 10^{-3}$
& 0&26 & $4\ 10^{-2}$ & 0&047 & $10\ 10^{-3}$ $^{\mathrm{ a}}$ & 0&047 & $7\
10^{-3}$ $^{\mathrm{ a}}$ & $38\pm8$ \\

Spitzer GFLS (105.6,+16) & 0&072 & $9\ 10^{-3}$ & 0&046 & $5\ 10^{-3}$
& 0&34 & $4\ 10^{-2}$ & 0&028 & $5\ 10^{-3}$ $^{\mathrm{ a}}$ & 0&044 & $5\
10^{-3}$ $^{\mathrm{ a}}$ & $52\pm12$ \\

Spitzer GFLS (254.4,+0) & 0&068 & $8\ 10^{-3}$ & 0&043 & $5\ 10^{-3}$
& 0&37 & $5\ 10^{-2}$ & 0&027 & $5\ 10^{-3}$ $^{\mathrm{ a}}$ & 0&041 & $5\
10^{-3}$ $^{\mathrm{ a}}$ & $52\pm12$ \\

\hline

Spitzer GLIMPSE (27.5) & 0&065 & $8\ 10^{-3}$ & 0&052 & $7\ 10^{-3}$ &
0&32 & $4\ 10^{-2}$ & 0&016 & $2.10^{-3}$ $^{\mathrm{ b}}$ & 0&049 & $6\
10^{-3}$ $^{\mathrm{ a}}$ & $60\pm9$ \\

\hline

\citet{Giard1994} & \multicolumn{3}{c}{-} & \multicolumn{3}{c}{-} &
\multicolumn{3}{c}{-} & 0&016 & $2.10^{-3}$ & \multicolumn{3}{c}{-} & \\

\citet{Tanaka1996} & \multicolumn{3}{c}{-} & \multicolumn{3}{c}{-} &
\multicolumn{3}{c}{-} & 0&016 & $4.10^{-3}$ & \multicolumn{3}{c}{-} & \\

\citet{Dwek1997} & \multicolumn{3}{c}{0.0306} &
\multicolumn{3}{c}{0.0339} & \multicolumn{3}{c}{-} &
\multicolumn{3}{c}{-} & \multicolumn{3}{c}{-} & \\

\citet{Arendt1998} & \multicolumn{3}{c}{0.0334} &
\multicolumn{3}{c}{0.0394} & \multicolumn{3}{c}{-} &
\multicolumn{3}{c}{-} & \multicolumn{3}{c}{-} & \\

\citet{Li2001} & \multicolumn{3}{c}{0.0477} &
\multicolumn{3}{c}{0.0224} & \multicolumn{3}{c}{0.278} &
\multicolumn{3}{c}{0.0264 $^{\mathrm{ a}}$} &
\multicolumn{3}{c}{0.0213 $^{\mathrm{ a}}$} & \\

\hline
\end{tabular}
\begin{list}{}{}
\item[$^\mathrm{a}$] derived from Eq.~\ref{eq:1} and \ref{eq:2}
\item[$^\mathrm{b}$] derived from \citet{Giard1994}
\end{list}
\end{table*}

Resulting IRAC colors are given in the first three columns of Table
\ref{tab:rdc}. For the GLIMPSE field, they are plotted, together with
the CVF spectrum on Fig.~\ref{fig:spectre}. The $R_{5.8/8.0}$ ratios,
given by the CVF spectrum (0.30) on the one hand, and by the GLIMPSE
field (0.32) on the other hand, are in a good agreement.

The measurements uncertainties come from the linear fitting process,
including gradient correction, and do not exceed 3\% on most of the
IRAC colors. Extinction correction increase these uncertainties up to
10\% whereas ionized gas corrections induce a negligible
uncertainty. Taking into account the photometric accuracy (see section
\ref{irac}), the final uncertainty is about 13\% on most of
$R_{3.6/8.0}$, $R_{4.5/8.0}$ and $R_{5.8/8.0}$, and reaches 16\% for
the GFLS (105.6,+8) field.

Our colors are averaged over large areas and do not give an account of
the small scale dispersion across the IRAC fields. However, they
already show strong variations from one field to another, especially
$R_{3.6/8.0}$ and $R_{4.5/8.0}$, whereas $R_{5.8/8.0}$ does not vary
that much around $0.3$.

\section{3.3 $\mu$m feature and continuum}
\label{sec:spectro}

For the GLIMPSE field, AROME observations are combined with the IRAC
colors to provide a spectrum of the diffuse emission from 3 to 5
$\mu$m. We generalize this derivation to the other fields.

\subsection{Inner Galaxy spectrum}

Based on reflection nebulae observations, we interpret the GLIMPSE
$R_{3.6/8.0}$ and $R_{4.5/8.0}$ colors with a PAH feature at 3.3
$\mu$m and an underlying continuum. Spectroscopic observations of
\object{NGC7023} suggest that the feature and the continuum both
contribute to the flux in IRAC 3.6 $\mu$m channel, whereas the flux in
IRAC 4.5 $\mu$m channel is dominated by the continuum
\citep{Sellgren1983}. The 3.3 $\mu$m feature has been
spectroscopically detected by IRTS in the inner Galaxy
\citep{Tanaka1996}. It has been measured photometrically by the AROME
experiment \citep{Giard1994} and \citet{Bernard1994} show that the
feature alone cannot account for the ISM emission in the DIRBE NIR
channels.

In reflection nebulae, the continuum is well described by a gray-body
with a color temperature $T_C = 1000-1500\ \mathrm{K}$. This
temperature is constrained thanks to photometric measurements on both
sides of the 3.3 $\mu$m feature. To interpret the GLIMPSE colors, we
construct a NIR diffuse emission spectrum with the spectral shape of
the feature from the ISO/SWS spectrum of \object{NGC7023} and a
gray-body continuum. The feature intensity is taken from the AROME
measurement, taking into account the spectral response of their
filters. We then fit the color temperature and intensity of the
continuum to match the IRAC $R_{3.6/8.0}$ and $R_{4.5/8.0}$
colors. The resulting spectrum is shown in Fig.~\ref{fig:spectre}. The
data does not constrain the spectral shape of the continuum. The
combination of IRAC colors and the AROME feature only determine the
ratio between the continuum emission in the 3.6 and 4.5 $\mu$m IRAC
channels. The color temperature of the gray body is $T_C=1100 \pm 300\
\mathrm{K} $, as computed from IRAC colors and AROME measurement. This
value is in the range given by \citet{Sellgren1983} for reflection
nebulae, where the physical conditions are much different from the
diffuse medium. The uncertainty on the color temperature is large
because we only have two measurements to determine it.

We can thus separate, for the inner Galaxy spectrum, the contributions
of the 3.3 $\mu$m feature and the continuum to the IRAC 3.6 $\mu$m
channel, which can be expressed as linear combinations of the
$R_{3.6/8.0}$ and $R_{4.5/8.0}$ colors (see Eq.~\ref{eq:1} and
\ref{eq:2}). We define the contribution of the feature as the ratio
between the flux of the 3.3 $\mu$m feature through the IRAC 3.6 $\mu$m
channel and the flux of the PAH emission through the IRAC 8.0 $\mu$m
channel, hereafter $R_{3.6/8.0,feat}$.

\begin{equation}
R_{3.6/8.0,feat} = R_{3.6/8.0} - 0.95\ \times R_{4.5/8.0}
\label{eq:1}
\end{equation}

The contribution of the continuum to the IRAC 3.6 $\mu$m channel, is
given by :

\begin{eqnarray}
R_{3.6/8.0,cont} &=& R_{3.6/8.0} - R_{3.6/8.0,feat} \nonumber \\
&=& 0.95 \times R_{4.5/8.0}
\label{eq:2}
\end{eqnarray}

The corresponding values are listed in Table \ref{tab:rdc}.

\subsection{Additional fields}

For the GFLS fields, we cannot determine the value of the color
temperature due to the lack of 3.3 $\mu$m feature measurements. Since
the ratio between the continuum flux in IRAC 3.6 and 4.5 $\mu$m
channels does not depend much on the color temperature, we assume that
it does not vary among IRAC fields. Within this assumption, we use
Eq.~\ref{eq:1} and \ref{eq:2}, and give the corresponding values of
$R_{3.6/8.0,feat}$ and $R_{3.6/8.0,cont}$ for the IRAC fields in the
two last columns of Table \ref{tab:rdc}. It appears that the field to
field variations of the 3.3 $\mu$m feature intensity are
important. There is a factor of $3.3$ between the smallest values, at
($l,b$) = (105.6,+0.3) and (27.5,+0), and the highest value found for
($l,b$) = (105.6,+8). On the contrary, the field to field variations
of the continuum contribution $R_{3.6/8.0,cont}$ are very weak. The
mean value is about $0.04$ and there is less than a factor of $1.8$
between the extrema values. In all fields, the continuum has a strong
contribution to the IRAC 3.6 $\mu$m channel. It accounts for one half
of the flux in IRAC 3.6 $\mu$m channel at ($l,b$) = (105.6,+8) and for
more than 75\% at ($l,b$) = (27.5,+0) and (105.6,+0.3). There is no
correlation between 3.3 $\mu$m feature and continuum colors.

\subsection{Comparison with DIRBE}

With the inner Galaxy spectrum (Fig.~\ref{fig:spectre}), we can
compare our IRAC colors to those obtained with DIRBE
\citep{Dwek1997,Arendt1998}. We obtain these numbers, given in Table
\ref{tab:rdc}, by convolving the Galactic spectrum with the
transmission curves of the instrument. For further studies, we give
the conversion factors in Table \ref{tab:conv}. Our colors are
significantly but slightly different from the DIRBE colors, especially
for $R_{3.6/8.0}$. However, this comparison does not lead to a unique
conclusion. DIRBE colors corresponds to high-latitude emission which
may well differ from those measured on IRAC fields. The difference may
also reflect systematic uncertainties in the DIRBE analysis associated
with stellar subtraction. Last, but not least, our spectroscopic model
may also contribute to a significant part of the difference.

\begin{table}[b]
\caption{Conversion factors from DIRBE or IRAS to IRAC fluxes, deduced
  from the Galactic spectrum.}
\label{tab:conv}
\centering
\begin{tabular}{c c}
\hline \hline
Instrument and channel & Conversion factor \\
\hline
$DIRBE_{3.5}/IRAC_{3.6}$ & 0.917 \\
$DIRBE_{4.9}/IRAC_{4.5}$ & 1.03 \\
$DIRBE_{12}/IRAC_{8.0}$ & 0.749 \\
$IRAS_{12}/IRAC_{8.0}$& 0.863 \\
\hline
\end{tabular}
\end{table}

\section{Spectral diagnostics of PAH size and ionization state}
\label{sec:pahs}

The PAH emission spectrum depends on both their charge state and size
distribution. \citet{Li2001} computed the diffuse ISM PAH spectrum
with a model based on laboratory data. In Table \ref{tab:rdc} we
compare our colors to those expected from their model. Their model
values are reasonably close to the colors which do not include
significant continuum contribution. We have developed our own model,
for several reasons. (1) The values listed by \citet{Li2001}
corresponds to a mixture between the Cold Neutral Medium (CNM), the
Warm Neutral Medium (WNM) and the Warm Ionized Medium (WIM) in
proportions 43\%, 43\% and 14\% in mass, which are characteristic of
high-latitude line of sight but do not apply to the low latitudes IRAC
fields for which a significant fraction of the gas is molecular and
thus are mostly sampling the CNM. The ionization state of PAHs along
these lines of sight might thus differ from that in \citet{Li2001}
model. (2) As pointed out in previous sections, the 3.3 $\mu$m
intensity measured through the $R_{3.6/8.0,feat}$ color varies by a
factor of 3 from field to field. We need to run a model over a grid of
parameters to translate these variations in terms of PAH mean size and
ionization state (see appendix \ref{sec:model} for details). (3) In a
recent study, \citet{Rapacioli2005} proposed a spectral decomposition
of \object{NGC7023} ISOCAM/CVF spectro-imaging data leading to
distinct emission spectra for cation and neutral PAHs. This opens the
possibility to define the PAH properties from observations rather than
laboratory measurements. Our model, an update of \citet{Desert1990},
is based on this spectral decomposition, from which we derive the PAH
cations and neutrals cross-sections, extending the work of
\citet{Rapacioli2005}.

We run the model, coupled with a module that computes the PAH
ionization fraction as a function of the PAH size. We apply this for
various PAH size distributions and values of the ionization parameter
$G\sqrt{T}/n_\mathrm{e}$, where $G$ is the integrated far ultraviolet
(6-13.6 eV) radiation field expressed in units of the Habing radiation
field, $T$ is the gas temperature and $n_\mathrm{e}$ is the electronic
density. The PAH size distribution of \citet{Desert1990} has a mean
size of 6 \AA\ or 45 carbon atoms, according to the relation
$a=0.9\sqrt{N_\mathrm{C}}$ between the PAH size $a$ in angstroms and
the number of carbon atoms $N_\mathrm{C}$. We vary the PAH mean size
by changing the exponent of the PAH power law size distribution
(standard value is $-3$) keeping fixed the values of the minimum and
maximum PAH sizes ($N_\mathrm{C}=20-180$ or $a=4-12\ $\AA).

We use two spectroscopic diagnostics to constrain the PAH properties :
the ratio between the flux in the band at 7.7 and the band at 11.3
$\mu$m (hereafter $R_{7.7/11.3}$) as well as $R_{3.6/8.0,feat}$.

\subsection{$R_{7.7/11.3}$ as a tracer of PAH ionization state}

\begin{figure}[t]
\resizebox{\hsize}{!}{\includegraphics{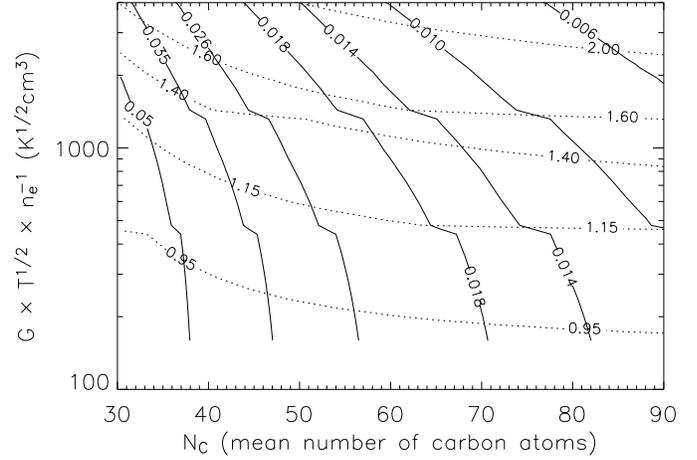}}
\caption{$R_{3.6/8.0,feat}$ ({\it solid lines}) and $R_{7.7/11.3}$
  ({\it dashed lines}) deduced from the model as a function of the PAH
  mean size and $G\sqrt{T}/n_\mathrm{e}$.}
\label{fig:cont}
\end{figure}

Using a Lorentzian decomposition that fits the PAH features between
6.2 and 12.7 $\mu$m, we calculate the ratio between the fluxes in the
band at 7.7 and the band at 11.3 $\mu$m (hereafter $R_{7.7/11.3}$) for
each PAH model spectrum. The central wavelengths of the Lorentzian
features are given as fixed inputs of the fitting process at 6.2, 7.6,
8.6, 11.3 and 12.7 $\mu$m, whereas the widths and amplitudes are set
free. Fig.~\ref{fig:cont} shows $R_{7.7/11.3}$ as a function of
$G\sqrt{T}/n_\mathrm{e}$ and the PAH mean size. In the model and
within the range of values we consider, this ratio depends much more
on the PAH ionization than on the average size : neutral PAHs present
a lower $R_{7.7/11.3}$ ratio than ionized PAHs
\citep{Draine2001,Bakes2001a}. For a mean size of $N_\mathrm{C} = 45$
carbon atoms, $R_{7.7/11.3}$ grows from $0.95$ at
$G\sqrt{T}/n_\mathrm{e} = 250\ \mathrm{K}^{1/2}\ \mathrm{cm}^{3}$ to
$1.95$ at $G\sqrt{T}/n_\mathrm{e} = 4000\ \mathrm{K}^{1/2}\
\mathrm{cm}^{3}$, whereas it goes from $1.0$ to $2.2$ for a mean size
of $N_\mathrm{C} = 85$ carbon atoms, between the same values of
$G\sqrt{T}/n_\mathrm{e}$. This ratio traces the PAH ionization state.

\subsection{$R_{3.3/8.0,feat}$ as a tracer of PAH mean size}

Since the model does not include the continuum underlying the 3.3
$\mu$m feature, we immediately obtain $R_{3.6/8.0,feat}$ by dividing
the flux of the PAH model spectrum in the IRAC 3.6 $\mu$m channel by
its equivalent in the IRAC 8.0 $\mu$m channel. Fig.~\ref{fig:cont}
shows $R_{3.6/8.0,feat}$ as a function of the PAH mean size and
$G\sqrt{T}/n_\mathrm{e}$. Unlike $R_{7.7/11.3}$, $R_{3.6/8.0,feat}$
depends on the PAH size as well as their ionization, and it is much
more dependent on the PAH mean size when the PAHs are small. The 3.3
$\mu$m feature becomes fainter when the PAHs are big or ionized
\citep{Draine2001,Bakes2001a}. For a mean size of $N_\mathrm{C} = 45$
carbon atoms, $R_{3.6/8.0,feat}$ goes from $0.0175$ when
$G\sqrt{T}/n_\mathrm{e} = 4000\ \mathrm{K}^{1/2}\ \mathrm{cm}^{3}$ to
$0.0375$ when $G\sqrt{T}/n_\mathrm{e} = 250\ \mathrm{K}^{1/2}\
\mathrm{cm}^{3}$ , and from $0.005$ to $0.0125$ for $N_\mathrm{C} =
85$ between the same values of $G\sqrt{T}/n_\mathrm{e}$. This ratio,
when coupled with the previous one, constrains the PAH mean size.

\section{PAH size and ionization state across the diffuse ISM}

\begin{figure}[t]
\resizebox{\hsize}{!}{\includegraphics{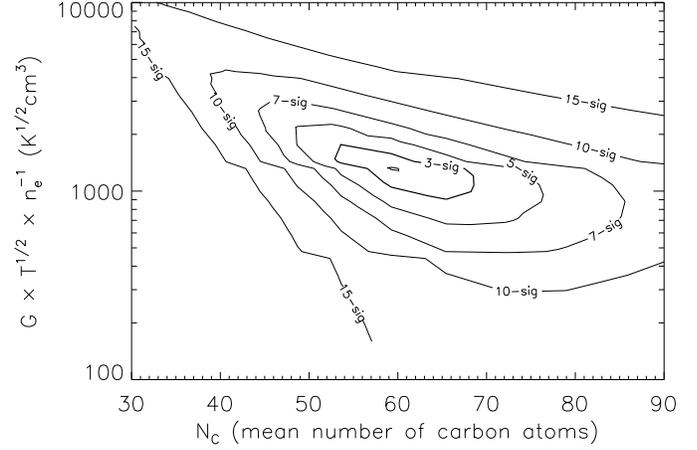}}
\caption{Iso-$\chi^2$ contours deduced from a fitting process of
  $R_{3.6/8.0}$, $R_{4.5/8.0}$, $R_{5.8/8.0}$, $R_{7.7/11.3}$ and
  $R_{3.6/8.0,feat}$, plotted as a function of $N_\mathrm{C}$ and
  $G\sqrt{T}/n_\mathrm{e}$. For each couple of $N_\mathrm{C}$ and
  $G\sqrt{T}/n_\mathrm{e}$, we take the lowest $\chi^2$, depending on
  the intensity and color temperature of the continuum. Only the
  statistical dispersion on the colors is taken into account on this
  figure. Systematic and statistical dispersions account for the same
  in the resulting uncertainties on the parameters.}
\label{fig:chi2}
\end{figure}

Combining $R_{7.7/11.3}$ and $R_{3.6/8.0,feat}$ (see
Fig.~\ref{fig:cont}), we can now constrain the PAH mean size and
ionization state for our inner Galactic spectrum, for which we have
both these measurements. Adding the other IRAC colors $R_{3.6/8.0}$,
$R_{4.5/8.0}$ and $R_{5.8/8.0}$, we simultaneously constrain the shape
and strength of the continuum. We also discuss the variations of the
PAH properties across the Galaxy.

\subsection{Inner Galactic diffuse medium}

\begin{figure}[t]
\resizebox{\hsize}{!}{\includegraphics{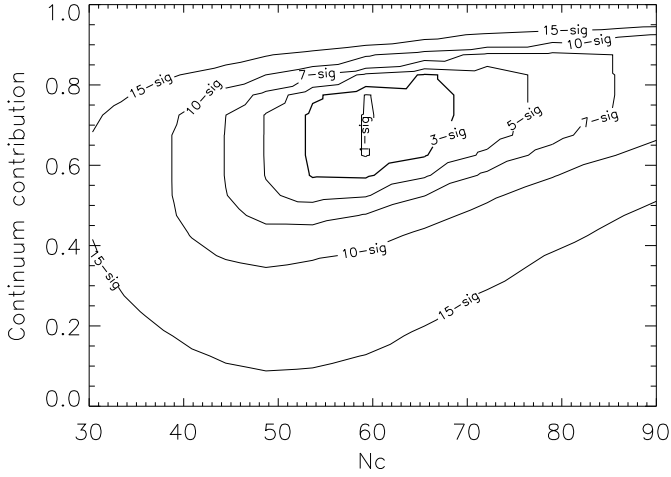}}
\caption{Iso-$\chi^2$ contours deduced from a fitting process of
  $R_{3.6/8.0}$, $R_{4.5/8.0}$, $R_{5.8/8.0}$, $R_{7.7/11.3}$ and
  $R_{3.6/8.0,feat}$ as a function of the continuum contribution to
  IRAC 3.6 $\mu$m channel and the PAH mean size. For each couple of
  $N_\mathrm{C}$ and continuum contribution, we take the lowest
  $\chi^2$, depending on $G\sqrt{T}/n_\mathrm{e}$ . Only the
  statistical dispersion on the colors is taken into account on this
  figure. Systematic and statistical dispersions account for the same
  in the resulting uncertainties on the parameters.}
\label{fig:chi2_cont}
\end{figure}

\begin{figure}[t]
\resizebox{\hsize}{!}{\includegraphics{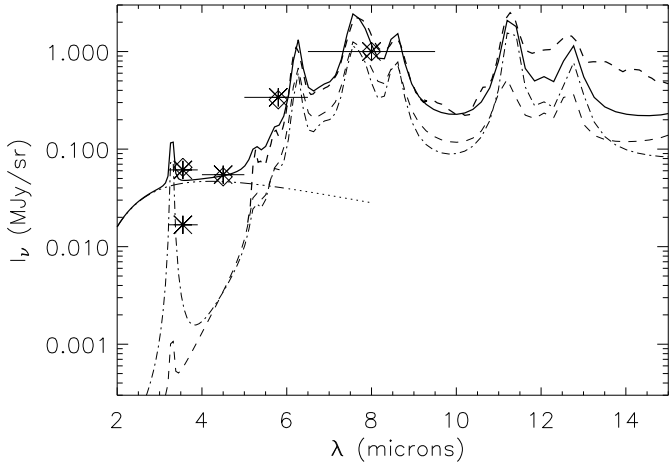}}
\caption{{\it Bold dashed line} : CVF spectrum of the diffuse Galactic
  emission (for $N_\mathrm{H}=10^{21} \mathrm{cm}^{-2}$) centered on
  the Galactic coordinates (26.8,+0.8). {\it Bold solid line} : best
  fit resulting from the model, adding three components : PAH cations
  ({\it dashed line}), PAH neutrals ({\it dash-dot line}), and a
  continuum ({\it dash-3-dot line}). The parameters used for this fit
  are listed in Table \ref{tab:sp_param}. The origin of the continuum
  is discussed in section \ref{sec:cont}.}
\label{fig:sp_chi2}
\end{figure}

For our inner Galactic spectrum, $R_{7.7/11.3} = 1.5\pm0.1$ and
$R_{3.6/8.0,feat} = 0.016\pm0.002$. We compute a best-fit process over
four parameters ($N_\mathrm{C}$, $G\sqrt{T}/n_\mathrm{e}$, and the
continuum intensity and color temperature) taking into account five
constraints (the two ratios previously detailes, as well as the three
IRAC colors $R_{3.6/8.0}$, $R_{4.5/8.0}$ and $R_{5.8/8.0}$ given in
Table \ref{tab:rdc}). The resulting $\chi^2$ is plotted on
Fig.~\ref{fig:chi2} and \ref{fig:chi2_cont}. The 3-$\sigma$ error bars
on the parameters correspond to $\chi^2 = 10$. Fig.~\ref{fig:sp_chi2}
shows the best-fit spectrum and corresponding physical parameters are
listed in Table \ref{tab:sp_param}.

A PAH mean size of $N_\mathrm{C} = 60$ carbon atoms is larger by a
factor of 1.3 than the mean size of \citet{Desert1990}. Along the
inner Galaxy line of sight, half of the gas is molecular (see section
\ref{sec:obs}). A consistency check on the model is provided by the
derived value of $G\sqrt{T}/n_\mathrm{e}$ which should be close to
that of the CNM. According to \citet{Li2001} and
\citet{Weingartner2001}, in a radiation field $G \simeq 1$,
$G\sqrt{T}/n_\mathrm{e} = 250-380\ \mathrm{K}^{1/2}\ \mathrm{cm}^{3}$
for the CNM, $2200-2900\ \mathrm{K}^{1/2}\ \mathrm{cm}^{3}$ for the
WNM and $1000\ \mathrm{K}^{1/2}\ \mathrm{cm}^{3}$ for the
WIM. Considering that the radiation field increases from the inner
Galaxy to the solar neighborhood by a factor of $\sim 3$
\citep{Sodroski1997}, the ionization parameter for the CNM, near the
molecular ring, is about $750-1140\ \mathrm{K}^{1/2}\
\mathrm{cm}^{3}$, whereas the WIM and WNM are about $3000\
\mathrm{K}^{1/2}\ \mathrm{cm}^{3}$ and $6600-8700\ \mathrm{K}^{1/2}\
\mathrm{cm}^{3}$. Our Galactic spectrum is then in a good agreement
with a CNM dominated medium, assuming that the mean CNM electron
density and gas temperature are constant across the Galaxy. From the
ionization parameter, we derive that the PAHs are half-neutral and
half-cation. The contribution of the continuum in IRAC 3.6 $\mu$m
channel, $70\%\pm12\%$ (see Fig.~\ref{fig:chi2_cont}), which
corresponds to $R_{3.6/4.5,cont} = 0.90\pm0.15$, is in agreement with
the one empirically determined in section \ref{sec:spectro}.

\begin{table}
\caption{Best fit model output. Uncertainties result from statistical
  and systematic dispersions, which account for the same in the
  resulting uncertainties on the parameters.}
\label{tab:sp_param}
\centering
\begin{tabular}{l r@{$\pm$}l}
\hline \hline
Model output & \multicolumn{2}{c}{Value} \\
\hline
$N_\mathrm{C}$ & 60&9\\
$G\sqrt{T}/n_\mathrm{e}$ & 1350&450$\ \mathrm{K}^{1/2}\ \mathrm{cm}^{3}$ \\
$PAH^+/PAH$ & 42\%&7\% \\
Continuum contribution to IRAC 3.6 & 70\%&12\% \\
\hline
\end{tabular}
\end{table}

\subsection{Field-to-field variations}
\label{sec:var}

\citet{Sakon2004} have measured 6.2, 7.7, 8.6 and 11.3 $\mu$m features
at different Galactic longitudes along the Galactic plane
with IRTS spectroscopic data. They found that the 8.6 and 11.3 $\mu$m
features were systematically stronger relative to the 6.2 and 7.7
$\mu$m features in the outer Galaxy than in the inner Galaxy. They
suggest that PAH dehydrogenation or compactness may explain such band
to band variations rather than ionization, even if they do not derive
definite conclusions. Our CVF data allow us to address this question.

In Table \ref{tab:r7_11}, we list $R_{7.7/11.3}$ values for the
different CVF lines of sight presented in section \ref{sec:obs}. The
Lorentzian decomposition is made difficult on the CVF spectra by the
weak S/N of some spectra and by the short range of covered wavelengths
below 6.2 and above 12.7 $\mu$m. In order to estimate the
uncertainties of this measure, we try different methods, that differ
by the lorentzians parameters we fix or set free (position, amplitude
or width). We give the ratios in Table \ref{tab:r7_11} where we also
list the \citet{Sakon2004} values. Our two values for the inner galaxy
(G26.8 and G30) are lower than those of \citet{Sakon2004} and there
are no signs of a systematic Galactic gradient in the CVF data
nor in $R_{3.6/8.0,feat}$ color (see Table \ref{tab:rdc}).

\begin{table}
\caption{$R_{7.7/11.3}$ from CVF lines of sight and from \citet{Sakon2004}.}
\label{tab:r7_11}
\centering
\begin{tabular}{l r r r@{.}l@{$\pm$}r@{.}l}
\hline \hline
Line of sight & \multicolumn{2}{c}{Galactic coordinates} & \multicolumn{4}{c}{$R_{7.7/11.3}$} \\
 &  \multicolumn{2}{c}{(l,b)} & \multicolumn{4}{c}{} \\
\hline
G26.8    & \multicolumn{2}{c}{(26.8,0.8)}    & 1&5  & 0&1 \\
G34.1    & \multicolumn{2}{c}{(34.1,13.4)}   & 1&6  & 0&2 \\
G30      & \multicolumn{2}{c}{(30.0,3.0)}    & 1&3  & 0&1 \\
G299.7   & \multicolumn{2}{c}{(299.7,-16.3)} & 1&2  & 0&3 \\
\hline
\citet{Sakon2004} l=-8   & \multicolumn{2}{c}{(-8,0)}   & 2&85 & 0&17 \\
\citet{Sakon2004} l=48   & \multicolumn{2}{c}{(48,0)}   & 1&99 & 0&05 \\
\citet{Sakon2004} l=-132 & \multicolumn{2}{c}{(-132,0)} & 1&42 & 0&40 \\
\citet{Sakon2004} l=172  & \multicolumn{2}{c}{(172,0)}  & 1&65 & 0&37 \\
\hline
\end{tabular}
\end{table}

Within our model, the observed variations in the $R_{3.6/8.0,feat}$
color of the Spitzer fields must trace variations in the PAH mean
size. We refer to the solid lines plotted on
Fig.~\ref{fig:cont}. Since the stellar radiation field decreases from
the inner to the outer galaxy, we consider that
$G\sqrt{T}/n_\mathrm{e}$ is lower for the GFLS fields than for the
GLIMPSE field (we assume that $T$ and $n_\mathrm{e}$ are almost
constant from field to field). Besides, we use the CNM value of
\citet{Li2001} and \citet{Weingartner2001} $G\sqrt{T}/n_\mathrm{e}
\simeq 300\ \mathrm{K}^{1/2}\ \mathrm{cm}^{3}$ as a lower limit
because some of the gas must be in the WNM and WIM phases, where this
ionization parameter is higher. This range of ionization factors
allows us to constrain the mean size for each field, using the value
of $R_{3.6/8.0,feat}$ color. The derived PAH mean sizes are given in
Table \ref{tab:rdc}. To estimate the error bars, we use
$R_{3.6/4.5,cont} = 0.9^{+0.1}_{-0.2}$ (resulting from the fit) in
Eq.~\ref{eq:1} and \ref{eq:2}. There is a factor of $\sim 2$ between
the two extreme values, at (105.6,+8) with $38\pm8$ carbon atoms and
at (105.6,+0.3) with $80\pm20$ carbon atoms. The other PAH mean sizes
are about $50-60$ carbon atoms. What clearly appears is that we
observe significant variations in the PAH mean size : the field at
(105.6,+8) has much smaller PAHs than the field at (105.6,+0.3) and
the GLIMPSE field.

\section{The origin of the NIR continuum}
\label{sec:cont}

The NIR dust continuum, first detected in visual reflection nebulae
\citep{Sellgren1983} and observed in galaxies \citep{Lu2003}, is found
to also exist in the diffuse ISM, where the radiation field is
thousand times lower than in a reflection nebula like
\object{NGC7023}. This continuum accounts for 50\% to 80\% of the IRAC
3.6 $\mu$m channel intensity in the diffuse medium and its
field-to-field variations are weak, relative to the IRAC 8.0 $\mu$m
flux (see Table \ref{tab:rdc} and Fig.~\ref{fig:chi2_cont}). We
compare our values to the one found for galaxies by
\citet{Lu2003,Lu2004}. We estimate the contribution of free-free and
gas lines emission to their IRAC 3.6 and 4.5 $\mu$m colors, using the
$\mathrm{H_\alpha}$ to PAH emission ratio found by
\citet{Roussel2001}. The \citet{Lu2004} corrected colors $R_{3.6/8.0}
= 0.039 - 0.067$ and $R_{4.5/8.0} = 0.028 - 0.054$ are thus in
agreement with those we measure, even if they seem to be slightly
smaller.

The continuum observations raise two questions : what are the carriers
? what is the emission process ?

It is not scattered light. With the scattering properties for dust in
the diffuse ISM given by \citet{Li2001} and the NIR interstellar
radiation field derived from DIRBE NIR sky maps, we estimate the
intensity of the scattered light per H nucleon. The scattered light
accounts for 4\% and 1\% of the measured continuum in IRAC 3.6 and 4.5
$\mu$m channels in the inner Galactic spectrum.

PAH fluorescence has been proposed to account for the continuum. We
estimate the necessary photon conversion efficiency by dividing the
number of photons emitted in the continuum between 2.5 and 5 $\mu$m by
the number of photons absorbed in UV by PAHs. We assume that all of
the UV energy absorbed by PAHs is re-emitted in the infrared and the
mean energy of a UV photon absorbed by a PAH is $5.2 eV$
\citep{Li2001}. We thus find that the necessary photon conversion
efficiency is about 120\%. \citet{Gordon1998} performed a similar
calculation for the extended red emission (ERE) and obtained an
efficiency lower-limit of 10\% assuming that all the photons absorbed
by the dust are absorbed by the ERE producing material. Taking into
account that PAHs are responsible for one quarter of the dust UV
energy absorption, we find a NIR efficiency a factor of 3 higher than
the ERE efficiency. If we consider a solid photoluminescence process
-- due to VSGs -- instead of the molecular fluorescence, it leads to
an equivalent photon conversion efficiency, because VSGs absorb almost
the same energy as PAHs. In terms of energy, the molecular
fluorescence or solid photoluminescence have an efficiency of about
8.5\%.

\citet{Sellgren1984} have suggested that tri-dimensional grains of 45
to 100 carbon atoms that undergo stochastic heating like PAHs may be
the carriers of the continuum in reflection nebulae. As mentioned in
section \ref{sec:var}, the PAH mean size varies from $N_{\mathrm{C}} =
38\pm8$ carbon atoms to $N_{\mathrm{C}} = 80\pm20$ carbon atoms in the
different IRAC fields we analyze. However, in our analysis, the
continuum is not correlated with the PAH mean size. \citet{An2003}
found a systematic increase of the feature to continuum ratio with
increasing distance to the exciting star of \object{NGC7023}. This
could only reflect the dependence of the feature emission on the
ionization state and not tell us anything about the continuum.

The continuum emission questions the 3.3 $\mu$m feature
interpretation. If there is an efficient fluorescence mechanism for
the continuum, it could also account for the feature. Even if the
carriers of the continuum and feature are distinct, we cannot discard
the possibility that the emission process is the same. One cannot be
secure about the feature interpretation as long as the continuum
origin remains unclear.

\section{Conclusion}

We combine IRAC (GFLS and GLIMPSE fields) and ISOCAM/CVF data to
characterize the Near to Mid-IR Galactic diffuse ISM emission. Our
results are as follows :

\begin{itemize}
\item Extended diffuse emission is visible in most fields at the
  four IRAC wavelengths. We obtain IRAC colors by correlating the
  spatial distribution of the extended emission from one channel to
  another. CVF spectroscopic data available on our main field,
  pointing towards the inner Galaxy, directly demonstrates that the
  emission in the 5.8 and 8.0 $\mu$m channels comes from PAHs (mainly
  6.2 and 7.7 $\mu$m features). On the same field, comparison with 3.3
  $\mu$m feature photometric measurements show that PAHs accounts for
  only 25\% of the IRAC 3.6 $\mu$m channel flux. A NIR continuum,
  previously seen in reflection nebulae, must be present to account
  for the remaining fraction of the IRAC 3.6 $\mu$m channel flux and
  the totality of the IRAC 4.5 $\mu$m channel flux. This decomposition
  is generalized to the other fields to obtain the 3.3 $\mu$m feature
  and continuum contributions to the IRAC 3.6 $\mu$m channel flux.

\item Among the fields we analyze, the PAH colors exhibit significant
  variations. On the contrary, the continuum remains almost constant,
  relative to the PAH flux in the IRAC 8.0 $\mu$m channel and is in
  agreement with the values found for galaxies.

\item We interpret observed colors and their variations in terms of
  PAH mean size and ionization state. For this, we update the model of
  \citet{Desert1990} introducing different IR cross-sections for
  cation and neutral PAHs, determined from \object{NGC7023} IRAC and
  ISOCAM/CVF observations. We use two spectroscopic diagnostics within
  the model and fit our main field observations with a set of 5
  colors. The ratio between the 7.7 and the 11.3 $\mu$m features leads
  to a constraint on $G\sqrt{T}/n_\mathrm{e}$ which governs the
  equilibrium between cationic and neutral PAHs. For the GLIMPSE
  field, we find $G\sqrt{T}/n_\mathrm{e} = 1300\pm200\
  \mathrm{K}^{1/2}\ \mathrm{cm}^{3}$ in agreement with theoretical
  expectations for a CNM dominated line of sight and the molecular
  ring radiation field. The cation fraction is about 50\%. We also use
  the ratio between the 3.3 $\mu$m feature seen by IRAC 3.6 $\mu$m
  channel and the PAH emission in IRAC 8.0 $\mu$m channel to derive
  constraints on the PAH mean size. Combining both ratios, we show
  that there is a significant dispersion in the PAH mean size, which
  varies among the IRAC fields, from 38 to 80 carbon atoms.

\item The continuum intensity is not correlated with the 3.3 $\mu$m
  feature, which implies that the continuum carriers might not be
  PAHs. It is not scattered light. A photon conversion efficiency of
  about 120\% (energy conversion of 8.5\%) is necessary to account for
  it by PAH fluorescence or VSG photoluminescence.
\end{itemize}

The IRAC colors we have measured should be useful for further analysis
of the diffuse emission in IRAC images. An interesting development of
this work should be the analysis of the IRAC colors and their
variations on small spatial scales towards nearby clouds at high
Galactic latitude to probe the evolution of PAH size distribution in
interstellar clouds. If variations of PAH size distribution is
confirmed, it could be correlated with variations in the extinction
curve and diffuse infrared bands (DIBs) in order to trace the
contribution of small PAHs to these signatures.

\appendix
\section{Determination of neutral and cation PAH cross-sections}
\label{sec:model}

In order to interpret IRAC colors and their variations in terms of PAH
mean size and ionization state, we update the model of
\citet{Desert1990} to take into account the size depending ionization
state of the PAHs. We thus introduce distinct emission properties for
neutral and cationic PAHs. Immediately note that this model only gives
an account of the PAH features emission but not of the continuum
emission. The 3 to 13 $\mu$m absorption cross sections of these two
PAH forms are deduced from spectro-imaging CVF and IRAC observations
of \object{NGC7023}.

\begin{figure*}[!t]
\centering
\includegraphics[width=17cm]{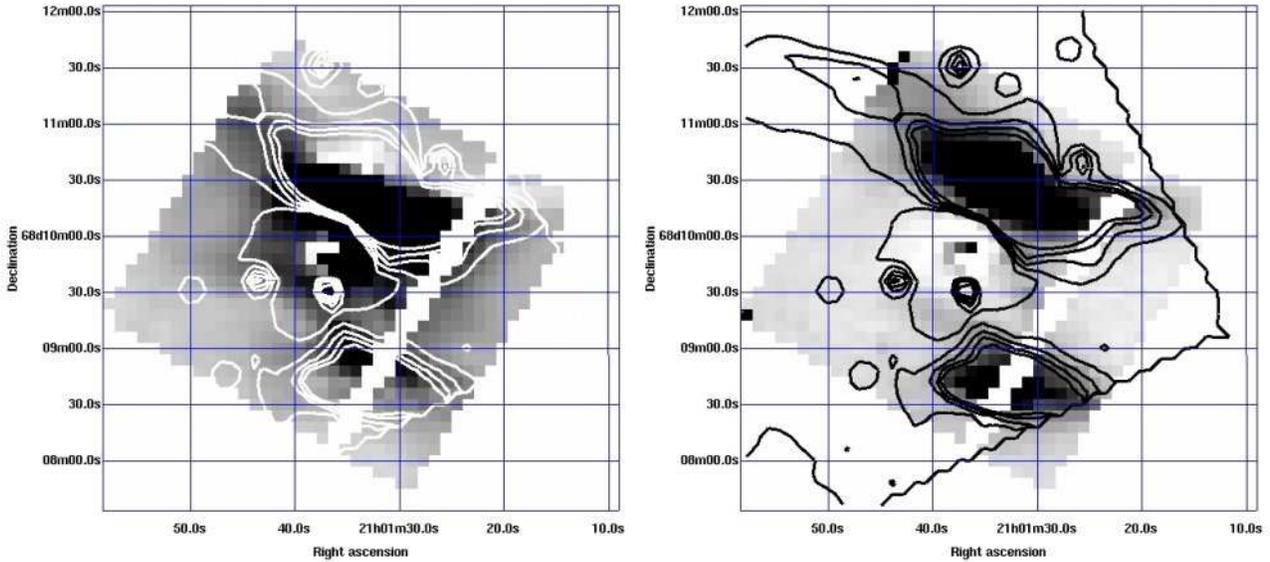}
\caption{\object{NGC7023} weight maps of PAH cations (\textit{left})
  and neutrals (\textit{right}), where the illuminating star is at the
  center and the black pixels code for the highest weights. The
  contours of $I_{\nu,3.3 feature}(i,j)$ are added in white
  (\textit{left}) or black (\textit{right}). Small scale contours
  (e.g. at ($\alpha,\delta$) = (21h01mn37s,
  68$\degr$11$\arcmin$30$\arcsec$)) are due to bright point sources,
  not taken into account in the fitting process. }
\label{fig:map}
\end{figure*}

First, emission spectra $I_{\nu}^{+}(\lambda)$ and
$I_{\nu}^{0}(\lambda)$ of pure PAH cations or neutrals are obtained
through the following procedure. We start from the
\citet{Rapacioli2005} spectra for $\lambda > 5 \mu$m based on their
linear decomposition of the \object{NGC7023} ISOCAM/CVF
observations. With these spectra, we make a linear decomposition of
CVF map which gives us weight-maps $CAT(i,j)$ and $NEU(i,j)$ of the
PAH neutrals and cations spectra contributions (see Eq.~\ref{eq:3} and
Fig.~\ref{fig:map}). $CONST(i,j,\lambda)$ accounts for emission of
larger dust particles (VSGs).

\begin{eqnarray}
I_{\nu}(i,j,\lambda) & = & I_{\nu}^{0}(\lambda)\times NEU(i,j) \nonumber\\
& & +\ I_{\nu}^{+}(\lambda)\times CAT(i,j) + CONST(i,j,\lambda)
\label{eq:3}
\end{eqnarray}

With these two weight-maps, we decompose the difference map between
IRAC 3.6 $\mu$m and IRAC 4.5 $\mu$m channel, according to
Eq.~\ref{eq:1}, to obtain $I_{\nu}^{+}(\lambda)$ and
$I_{\nu}^{0}(\lambda)$ at $\lambda = 3.6\ \mu m$. We also perform this
decomposition on the IRAC 8.0 $\mu$m channel.

\begin{figure}
\resizebox{\hsize}{!}{\includegraphics{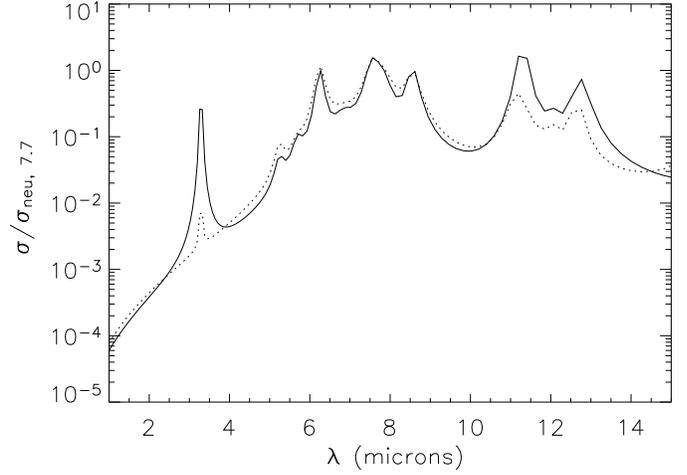}}
\caption{Neutral (solid lines) and cationic (dotted lines) PAH cross
  sections, for a 60 carbon atoms PAH, normalized to the 7.7 $\mu$m
  value for neutral PAH.}
\label{fig:pah}
\end{figure}

The second step is to invert the emission spectra into absorption
cross sections of PAH neutrals and cations. We assume that the PAH
size distribution is the one of \citet{Desert1990} model and we
approximate the spectral distribution of \object{NGC7023} radiation
field by a black body of $T_{eff} = 17000\ \mathrm{K}$ (the effective
temperature of the illuminating star). The inversion does not depend
on field intensity. We iterate on the cross sections values until the
model reproduces the emission spectra, and the intensity in IRAC 8.0
$\mu$m channel and 3.3 $\mu$m feature to an accuracy better than 10\%.

\begin{table}[!t]
\caption{Amplitudes and widths for PAH cations and neutrals that gives
  the integrated cross-sections, both normalized to the 3.3 $\mu$
  feature.}
\label{tab:section}
\centering
\begin{tabular}{r@{.}l r@{.}l c r@{.}l c}
\hline \hline
\multicolumn{2}{c}{$\lambda\ (\mu m)$} &
\multicolumn{2}{c}{$a_{neutral}$} & $W_{neutral}\ (cm^{-1})$ &
\multicolumn{2}{c}{$a_{cation}$} & $W_{cation}\ (cm^{-1})$ \\
\hline
3&3  &  1&0     & 20 &   1&0   & 20  \\
5&25 &  0&013   & 43 &   0&82  & 43  \\
5&7  &  0&026   & 43 &   0&69  & 43  \\
6&2  &  0&401   & 25 &   24&0  & 32  \\
6&85 &  0&064   & 65 &   3&65  & 85  \\
7&5  &  0&129   & 26 &   9&60  & 36  \\
7&6  &  0&450   & 31 &   17&3  & 26  \\
7&83 &  0&206   & 31 &   14&4  & 45  \\
8&6  &  1&29    & 20 &   54&7  & 25  \\
11&3 &  3&34    & 10 &   28&8  & 20  \\
12&0 &  0&154   & 12 &   5&28  & 15  \\
12&7 &  0&900   & 13 &   21&1  & 10  \\
\hline
\end{tabular}
\end{table}

The final amplitudes and widths for both PAH cations and neutrals are
given in Table \ref{tab:section}. The integrated cross-sections are
given by the products of these two quantities. The cross-sections
ratios for a given ionization state differ from those of
\citet{Li2001} by up to a factor of a few. The good correlation
between PAH neutrals and the feature clearly appears on
Fig.~\ref{fig:map}, where contours are the 3.3 $\mu$m feature
intensity, whereas PAH cations are closer to the star. The
cross-sections of pure PAH cations and neutrals, for the Mathis
radiation field and for the standard size distribution are shown in
Fig.~\ref{fig:pah}.

The model is coupled with a module that computes the cation, neutral
and anion fractions for each PAH size from the local balance between
the photo-ionization, the electron recombination, the electronic
attachment and the photo-detachment processes as quantified by
(\citet{LePage2001}). The cation to neutral fraction depends on
$G\sqrt{T}/n_\mathrm{e}$, where $G$ is the integrated far ultraviolet
(6-13.6 eV) radiation field expressed in units of the Habing radiation
field, $T$ is the gas temperature and $n_\mathrm{e}$ is the electronic
density. The fraction of PAH anions is computed to be small for
diffuse ISM physical conditions ($G\sqrt{T}/n_\mathrm{e}\sim
100-3000$). The small fraction of anions is assumed to have the same
emission properties than neutrals.

\bibliographystyle{aa}
\bibliography{3949flag.bib}

\begin{thebibliography}{39}
\expandafter\ifx\csname natexlab\endcsname\relax\def\natexlab#1{#1}\fi

\bibitem[{{Altenhoff} {et~al.}(1979){Altenhoff}, {Downes}, {Pauls}, \&
  {Schraml}}]{Altenhoff1979}
{Altenhoff}, W.~J., {Downes}, D., {Pauls}, T., \& {Schraml}, J. 1979, \aaps,
  35, 23

\bibitem[{{An} \& {Sellgren}(2003)}]{An2003}
{An}, J.~H. \& {Sellgren}, K. 2003, \apj, 599, 312

\bibitem[{{Arendt} {et~al.}(1998){Arendt}, {Odegard}, {Weiland}, {Sodroski},
  {Hauser}, {Dwek}, {Kelsall}, {Moseley}, {Silverberg}, {Leisawitz},
  {Mitchell}, {Reach}, \& {Wright}}]{Arendt1998}
{Arendt}, R.~G., {Odegard}, N., {Weiland}, J.~L., {et~al.} 1998, \apj, 508, 74

\bibitem[{{Bakes} {et~al.}(2001){Bakes}, {Tielens}, \&
  {Bauschlicher}}]{Bakes2001a}
{Bakes}, E.~L.~O., {Tielens}, A.~G.~G.~M., \& {Bauschlicher}, C.~W. 2001, \apj,
  556, 501

\bibitem[{{Beckert} {et~al.}(2000){Beckert}, {Duschl}, \&
  {Mezger}}]{Beckert2000}
{Beckert}, T., {Duschl}, W.~J., \& {Mezger}, P.~G. 2000, \aap, 356, 1149

\bibitem[{{Bernard} {et~al.}(1994){Bernard}, {Boulanger}, {Desert}, {Giard},
  {Helou}, \& {Puget}}]{Bernard1994}
{Bernard}, J.~P., {Boulanger}, F., {Desert}, F.~X., {et~al.} 1994, \aap, 291,
  L5

\bibitem[{{Boulanger} {et~al.}(2000){Boulanger}, {Abergel}, {Cesarsky},
  {Bernard}, {Miville Desch{\^ e}nes}, {Verstraete}, \&
  {Reach}}]{Boulanger2000}
{Boulanger}, F., {Abergel}, A., {Cesarsky}, D., {et~al.} 2000, in ESA SP-455:
  ISO Beyond Point Sources: Studies of Extended Infrared Emission, 91--+

\bibitem[{{Boulanger} {et~al.}(1985){Boulanger}, {Baud}, \& {van
  Albada}}]{Boulanger1985}
{Boulanger}, F., {Baud}, B., \& {van Albada}, G.~D. 1985, \aap, 144, L9

\bibitem[{{Boulanger} {et~al.}(2005){Boulanger}, {Lorente}, {Miville Desch{\^
  e}nes}, {Abergel}, {Blommaert}, {Cesarsky}, {Okumura}, {P{\' e}rault}, \&
  {Reach}}]{Boulanger2005}
{Boulanger}, F., {Lorente}, R., {Miville Desch{\^ e}nes}, M.~A., {et~al.} 2005,
  \aap, 436, 1151

\bibitem[{{Burton} \& {Hartmann}(1994)}]{Burton1994}
{Burton}, W.~B. \& {Hartmann}, D. 1994, \apss, 217, 189

\bibitem[{{Cohen} {et~al.}(1986){Cohen}, {Dame}, \& {Thaddeus}}]{Cohen1986}
{Cohen}, R.~S., {Dame}, T.~M., \& {Thaddeus}, P. 1986, \apjs, 60, 695

\bibitem[{{Desert} {et~al.}(1990){Desert}, {Boulanger}, \&
  {Puget}}]{Desert1990}
{Desert}, F.-X., {Boulanger}, F., \& {Puget}, J.~L. 1990, \aap, 237, 215

\bibitem[{{Dickinson} {et~al.}(2003){Dickinson}, {Davies}, \&
  {Davis}}]{Dickinson2003}
{Dickinson}, C., {Davies}, R.~D., \& {Davis}, R.~J. 2003, \mnras, 341, 369

\bibitem[{{Draine} \& {Li}(2001)}]{Draine2001}
{Draine}, B.~T. \& {Li}, A. 2001, \apj, 551, 807

\bibitem[{{Dwek} {et~al.}(1997){Dwek}, {Arendt}, {Fixsen}, {Sodroski},
  {Odegard}, {Weiland}, {Reach}, {Hauser}, {Kelsall}, {Moseley}, {Silverberg},
  {Shafer}, {Ballester}, {Bazell}, \& {Isaacman}}]{Dwek1997}
{Dwek}, E., {Arendt}, R.~G., {Fixsen}, D.~J., {et~al.} 1997, \apj, 475, 565

\bibitem[{{Giard} {et~al.}(1994){Giard}, {Lamarre}, {Pajot}, \&
  {Serra}}]{Giard1994}
{Giard}, M., {Lamarre}, J.~M., {Pajot}, F., \& {Serra}, G. 1994, \aap, 286, 203

\bibitem[{{Gordon} {et~al.}(1998){Gordon}, {Witt}, \& {Friedmann}}]{Gordon1998}
{Gordon}, K.~D., {Witt}, A.~N., \& {Friedmann}, B.~C. 1998, \apj, 498, 522

\bibitem[{{Hummer} \& {Storey}(1987)}]{Hummer1987}
{Hummer}, D.~G. \& {Storey}, P.~J. 1987, \mnras, 224, 801

\bibitem[{{Indebetouw} {et~al.}(2005){Indebetouw}, {Mathis}, {Babler}, {Meade},
  {Watson}, {Whitney}, {Wolff}, {Wolfire}, {Cohen}, {Bania}, {Benjamin},
  {Clemens}, {Dickey}, {Jackson}, {Kobulnicky}, {Marston}, {Mercer},
  {Stauffer}, {Stolovy}, \& {Churchwell}}]{Indebetouw2005}
{Indebetouw}, R., {Mathis}, J.~S., {Babler}, B.~L., {et~al.} 2005, \apj, 619,
  931

\bibitem[{{Le Page} {et~al.}(2001){Le Page}, {Snow}, \&
  {Bierbaum}}]{LePage2001}
{Le Page}, V., {Snow}, T.~P., \& {Bierbaum}, V.~M. 2001, \apjs, 132, 233

\bibitem[{{Leger} \& {Puget}(1984)}]{Leger1984}
{Leger}, A. \& {Puget}, J.~L. 1984, \aap, 137, L5

\bibitem[{{Li} \& {Draine}(2001)}]{Li2001}
{Li}, A. \& {Draine}, B.~T. 2001, \apj, 554, 778

\bibitem[{{Lu}(2004)}]{Lu2004}
{Lu}, N. 2004, \apjs, 154, 286

\bibitem[{{Lu} {et~al.}(2003){Lu}, {Helou}, {Werner}, {Dinerstein}, {Dale},
  {Silbermann}, {Malhotra}, {Beichman}, \& {Jarrett}}]{Lu2003}
{Lu}, N., {Helou}, G., {Werner}, M.~W., {et~al.} 2003, \apj, 588, 199

\bibitem[{{Mattila} {et~al.}(1996){Mattila}, {Lemke}, {Haikala}, {Laureijs},
  {Leger}, {Lehtinen}, {Leinert}, \& {Mezger}}]{Mattila1996}
{Mattila}, K., {Lemke}, D., {Haikala}, L.~K., {et~al.} 1996, \aap, 315, L353

\bibitem[{{Rapacioli} {et~al.}(2005){Rapacioli}, {Joblin}, \&
  {Boissel}}]{Rapacioli2005}
{Rapacioli}, M., {Joblin}, C., \& {Boissel}, P. 2005, \aap, 429, 193

\bibitem[{{Reynolds}(1992)}]{Reynolds1992}
{Reynolds}, R.~J. 1992, \apjl, 392, L35

\bibitem[{{Rieke} \& {Lebofsky}(1985)}]{Rieke1985}
{Rieke}, G.~H. \& {Lebofsky}, M.~J. 1985, \apj, 288, 618

\bibitem[{{Roussel} {et~al.}(2001){Roussel}, {Sauvage}, {Vigroux}, \&
  {Bosma}}]{Roussel2001}
{Roussel}, H., {Sauvage}, M., {Vigroux}, L., \& {Bosma}, A. 2001, \aap, 372,
  427

\bibitem[{{Sakon} {et~al.}(2004){Sakon}, {Onaka}, {Ishihara}, {Ootsubo},
  {Yamamura}, {Tanab{\' e}}, \& {Roellig}}]{Sakon2004}
{Sakon}, I., {Onaka}, T., {Ishihara}, D., {et~al.} 2004, \apj, 609, 203

\bibitem[{{Savage} \& {Mathis}(1979)}]{Savage1979}
{Savage}, B.~D. \& {Mathis}, J.~S. 1979, \araa, 17, 73

\bibitem[{{Schlegel} {et~al.}(1998){Schlegel}, {Finkbeiner}, \&
  {Davis}}]{Schlegel1998}
{Schlegel}, D.~J., {Finkbeiner}, D.~P., \& {Davis}, M. 1998, \apj, 500, 525

\bibitem[{{Sellgren}(1984)}]{Sellgren1984}
{Sellgren}, K. 1984, \apj, 277, 623

\bibitem[{{Sellgren} {et~al.}(1983){Sellgren}, {Werner}, \&
  {Dinerstein}}]{Sellgren1983}
{Sellgren}, K., {Werner}, M.~W., \& {Dinerstein}, H.~L. 1983, \apjl, 271, L13

\bibitem[{{Sodroski} {et~al.}(1997){Sodroski}, {Odegard}, {Arendt}, {Dwek},
  {Weiland}, {Hauser}, \& {Kelsall}}]{Sodroski1997}
{Sodroski}, T.~J., {Odegard}, N., {Arendt}, R.~G., {et~al.} 1997, \apj, 480,
  173

\bibitem[{{Tanaka} {et~al.}(1996){Tanaka}, {Matsumoto}, {Murakami}, {Kawada},
  {Noda}, \& {Matsuura}}]{Tanaka1996}
{Tanaka}, M., {Matsumoto}, T., {Murakami}, H., {et~al.} 1996, \pasj, 48, L53

\bibitem[{{van Diedenhoven} {et~al.}(2004){van Diedenhoven}, {Peeters}, {Van
  Kerckhoven}, {Hony}, {Hudgins}, {Allamandola}, \& {Tielens}}]{Van2004}
{van Diedenhoven}, B., {Peeters}, E., {Van Kerckhoven}, C., {et~al.} 2004,
  \apj, 611, 928

\bibitem[{{Verstraete} {et~al.}(2001){Verstraete}, {Pech}, {Moutou},
  {Sellgren}, {Wright}, {Giard}, {L{\' e}ger}, {Timmermann}, \&
  {Drapatz}}]{Verstraete2001}
{Verstraete}, L., {Pech}, C., {Moutou}, C., {et~al.} 2001, \aap, 372, 981

\bibitem[{{Weingartner} \& {Draine}(2001)}]{Weingartner2001}
{Weingartner}, J.~C. \& {Draine}, B.~T. 2001, \apjs, 134, 263

\end{thebibliography}

\end{document}